\documentclass[preprintnumbers,amsmath,amssymb,floatfix,10pt,prd,onecolumn,
superscriptaddress,nofootinbib]{revtex4}

\usepackage{amsfonts}
\usepackage{mathrsfs}
\usepackage{latexsym}
\usepackage{epsfig}
\usepackage{epstopdf}
\usepackage{graphicx}
\usepackage{amssymb}
\usepackage{amsmath}
\usepackage{dcolumn}
\usepackage{bm}
\usepackage{color}
\usepackage{xcolor}
\usepackage{comment}
\usepackage[cp1251]{inputenc}
\begin{document}

\title{\bf Impact of Tolman-Kuchowicz Potentials on Gauss-Bonnet Gravity and Isotropic Stellar Structures}

\author{Adnan Malik}
\email{adnan.malik@zjnu.edu.cn; adnan.malik@skt.umt.edu.pk: adnanmalik_chheena@yahoo.com}
\affiliation{School of Mathematical Sciences, Zhejiang Normal University, \\Jinhua, Zhejiang, China.}
\affiliation{Department of Mathematics, University of Management and Technology,\\ Sialkot Campus, Lahore, Pakistan.}

\begin{abstract}
\begin{center}
\textbf{Abstract}\\
\end{center}
\bigskip

In this manuscript, we study the behavior of charged isotropic compact stellar objects within the framework of modified $f(\mathcal{G})$ theory of gravity by considering the Tolman-Kuchowicz spacetime. We use some matching condition of spherically symmetric space-time with Bardeen’s model as an exterior geometry and examine the physical behavior of stellar structures. In the current analysis, we discuss the energy conditions to check the viability of our model. Many physical aspects have been examined, such as energy density, pressure evolution, equation of state parameter, and causality condition. Furthermore, an equilibrium condition can be visualized through the modified Tolman-Oppenheimer-Volkov equation. Further, we study mass-radius function, compactness and redshift function, which are some essential features of charged compact star model. It is worthwhile to mention here for the current study that our stellar structure in the background of Bardeen's model is more viable and stable.

\end{abstract}

\maketitle

\date{\today}
\section{Introduction}

The accelerated expansion of the universe is one of the most fascinating mysteries in cosmology, captivating scientists and enthusiasts alike. While conventional models, like General Relativity, have provided a solid foundation for understanding gravity, they struggle to fully explain this cosmic acceleration. This has led researchers to explore modified gravity theories, which offer alternative frameworks to reconcile observational data with theoretical predictions. These theories propose adjustments to the laws of gravity on cosmological scales, aiming to elucidate the mechanisms behind the universe's accelerated expansion without invoking concepts like dark energy. By introducing new fields or modifying existing equations, these theories present promising avenues for advancing our understanding of fundamental physics and the cosmos' evolution. Exploring the implications and observational signatures of modified gravity is essential for unraveling the mysteries of the universe while expanding the frontiers of scientific knowledge.  Koyama \cite{a1} examined recent developments in constructing modified gravity models as substitutes for dark energy and the advancement of cosmological tests for gravity.
Atazadeh and Sepangi \cite{a2} explored the Palatini formulation of modified gravity incorporating a Yukawa-like term, illustrating that this term accounts for the present exponential accelerated expansion of the universe and converges to the standard Friedmann cosmology under suitable conditions. Nojiri and Odintsov \cite{a3} developed the reconstruction program for the number of modified gravities: scalar-tensor theory, $f(R), F(G)$ and string-inspired, scalar-Gauss-Bonnet gravity and demonstrated that cosmological sequence of matter dominance, deceleration-acceleration transition and acceleration era may always emerge as cosmological solutions of such theory. Elizalde et al. \cite{a4} examined the proposed unified theory of exponential modified gravity, which accounts for early-time inflation and late-time acceleration, successfully navigating local tests and cosmological bounds, yet revealing instability during the inflationary era, with late-time evolution resembling that of the $\Lambda$CDM model for dark energy.
Granda \cite{a5} investigated unified inflation and late-time accelerated expansion, integrating exponential and $R^2$ corrections within modified gravity theory, while Dolgov and Kawasaki \cite{a6} demonstrated that recent proposals attributing cosmic acceleration to small curvature modification of gravity encounter severe instabilities and conflict with established gravitational interaction properties. Bahamonde along his collaborators \cite{a7} analysed a large number of cosmological models and show how the stability conditions allow them to be tightly constrained and even ruled out on purely theoretical grounds. Solanki et al., \cite{a8} analyzed the physical behavior of cosmological parameters such as density, deceleration, and the EoS parameters corresponding to the constrained values of the model parameters.

Modified $f(\mathcal{G})$ gravity presents an intriguing avenue for studying stellar structures due to its ability to capture gravitational effects beyond those described by General Relativity. This modified theory of gravity offers a more comprehensive framework for understanding the behavior of compact stellar objects. By incorporating additional terms based on the Gauss-Bonnet invariant $\mathcal{G}$, $f(\mathcal{G})$ gravity can potentially provide insights into phenomena such as gravitational collapse, neutron stars, and black holes. Exploring stellar structures within the context of $f(\mathcal{G})$ gravity allows researchers to investigate how modifications to gravitational laws influence the formation, stability, and evolution of these celestial bodies, contributing to our broader understanding of astrophysics and cosmology. Abbas et al. \cite{i7} explored the formation of anisotropic compact stars in modified Gauss-Bonnet, or $f(\mathcal{G})$ gravity, utilizing the Krori and Barua metric solutions to Einstein's field equations with anisotropic matter and power law model. Meanwhile, Ilyas et al. \cite{i8} delved into the interior configuration of static anisotropic spherical stellar charged structures within the framework of $f(\mathcal{G})$ gravity. Malik along with his collaborators \cite{i9} investigated the behavior of charge compact stars in the context of  $f(\mathcal{G})$ theory of gravity by using Bardeen geometry. Sharif and Saba \cite{i11} investigated anisotropic static spherically symmetric solutions in the framework of $f(\mathcal{G})$ gravity through gravitational decoupling approach. Shamir and Naz \cite{i10} explored the emergence of relativistic compact stellar objects in modified $f(\mathcal{G})$ gravity, employing the Noether symmetry approach to construct Noether symmetry generators and associated conserved quantities. De Felice and Tsujikawa \cite{i13} proposed an iterative technique to mitigate numerical instabilities linked to a substantial mass of the oscillating mode in $f(\mathcal{G})$ theory of gravity. Malik et al., \cite{a1} discussed the charged stellar structure in the background of the $f(\mathcal{G})$ theory of gravity by utilizing the Tolman-Kuchowicz spacetime. Naz et al., \cite{a2} investigated the Finch-Skea stellar structures obeying Karmarkar condition in modified $f(\mathcal{G})$ theory of gravity. Rashid et al., \cite{a3} used the conformal killing vectors to solve the Einstein–Maxwell field equations in the background of modified $f(\mathcal{G})$ gravity. Sharif and Ikram \cite{i14} explored the warm inflation in the background of $f(\mathcal{G})$ theory of gravity using scalar fields for the FRW universe model and constructed the field equations under slow-roll approximations and evaluate the slow-roll parameters, scalar and tensor power spectra and their corresponding spectral indices using viable power-law model. Garcia et al., \cite{i15} discussed the viability of $f(\mathcal{G})$ theory of gravity and presented the viability bounds arising from the energy conditions.

Tolman-Kuchowicz spacetime is of particular interest in the study of compact objects and gravitational physics because it provides a theoretical framework for describing the gravitational field around a spherically symmetric, non-rotating, and charged object. This spacetime solution allows for the exploration of the behavior of matter and curvature of spacetime in the vicinity of such compact objects. It is a significant model for understanding astrophysical phenomena like neutron stars and black holes, and it serves as a crucial tool for theoretical investigations into the properties and behavior of highly dense and massive celestial bodies. By examining Tolman-Kuchowicz spacetime, researchers can gain valuable insights into the complex interplay between matter, energy, and gravity in extreme astrophysical environments. Zubair et al. \cite{125e} examined the stellar spheres for the distribution of anisotropic fluid in the background of $f(T)$ modified gravity by using the Tolman-Kuchowicz space-time and Reissner Nordstram  geometry. Jasim et al., \cite{125f} studied a singularity-free model for the spherically symmetric anisotropic strange stars under Einstein's general theory of relativity by exploiting the Tolman-Kuchowicz metric. Biswas et al. \cite{125g} introduced a relativistic model for a static spherically symmetric anisotropic strange star employing Tolman-Kuchowicz metric potentials, while Bhar et al. \cite{125h} investigated anisotropic compact matter distributions in five-dimensional Einstein Gauss Bonnet gravity utilizing the same metric potentials. Recently, Malik along with his collaborators \cite{125i} conducted an extensive investigation of stellar compact structures within the frame of the newly proposed Ricci-Inverse theory of gravity and utilized the Tolman-Kuchowicz spacetime to explore these relativistic stars. Bhar et al., \cite{a9} obtained the analytically relativistic quintessence anisotropic spherical solutions in the $f(T)$ theory of gravity and imposed the pressure anisotropy condition by employing a metric potential of the Tolman–Kuchowicz (TK) type. Recently Bhar along with her collaborators \cite{a10} explored the relativistic anisotropic configuration of a hybrid star in the proximity of strange quark matter and normal baryonic matter within the context of $f (Q)$ theory of gravity. Some recent work on stellar structures and modified theories of gravities can be seen in \cite{a11,a12,a13,a14,a15,a16,a17,a18,a19}.

Motivated from the above literature, we investigate a detail analysis of isotropic stellar structures using Tolman-Kuchowicz metric potentials in Gauss-Bonnet gravity. To our knowledge, This is our first attempt to discuss the charged compact star with Tolman-Kuchowicz Space-time and Bardeen geometry in this theory of gravity. The paper is structured as follows: In Section II, we introduce demonstrations of a viable $f(\mathcal{G})$ gravity model. Section III explores matching conditions, while Section IV delves into the physical characteristics of charged compact stars, including energy density, radial pressure, tangential pressure, gradients thereof, energy conditions, stability criteria, adiabatic index, equation of state (EoS) parameters, mass function, compactness parameter, and redshift function. Finally, concluding remarks are presented in the last section.

\section{Isotropic matter distribution in $f(\mathcal{G})$ gravity}
The action for $f(\mathcal{G})$ gravity \cite{a4,a5,a6} is as follows
\begin{equation}\label{1}
\mathcal{S}= \frac{1}{k^2}\int d^{4}x
\sqrt{-g}\big[\frac{R}{2}+f(\mathcal{G})]+\mathcal{S}_{m}(g^{\mu\nu},\psi)+\mathcal{S}_{e}(g^{\mu\nu},\psi),
\end{equation}
where $R$, $\mathcal{G}$, and $\mathcal{S}_{m}$ denote the Ricci scalar, Guass-Bonnet invariant, and matter action respectively. Moreover,  $\mathcal{S}_{e}$ is lagrangian for electromagnetic field, which is defined as
\begin{equation}\label{1a}
\mathcal{S}_e=  -\frac{1}{16\pi}F_{\mu\nu}F_{\gamma\sigma}g^{\mu\gamma}g^{\nu\sigma},
\end{equation}
where $F_{\mu\nu}$ is electromagnetic field tensor, which is defined as
\begin{equation}\label{1b}
F^{\mu\nu}=\partial_{\nu}A_{\mu}-\partial_{\mu}A_{\nu},
\end{equation}
where $A_\mu$ is the electromagnetic four-potential. Einstein-Maxwell Field equations are defined as
\begin{equation}\label{1c}
F^{\mu\nu}_{;\nu}=-4 \pi j^{\mu},
\end{equation}
where $j^{\mu}=\sigma \nu^{\mu}$ is called electromagnetic four current vector and $\sigma$ represents as charge density. For our current analysis, we consider the static spherically symmetric space-time as
\begin{equation}\label{8}
ds^{2}=e^{\chi(r)}dt^{2}-e^{\xi(r)}dr^2-r^{2}(d\theta^{2}+\sin^{2}\theta d\phi^{2}).
\end{equation}

 The invariant quantity of Guass-Bonnet term is defined as
\begin{equation}\label{2}
\mathcal{G}=R^{2}-4R_{\mu\nu}R^{\mu\nu}+R_{\mu\nu\alpha\beta}R^{\mu\nu\alpha\beta},
\end{equation}
where
\begin{itemize}
\item $R_{\mu\nu}$ $\rightarrow$ Ricci tensor,
\item $R_{\mu\nu\alpha\beta}$ $\rightarrow$ Riemannian tensors.
\end{itemize}
Varying the action (\ref{1}) with respecting to $g_{\mu\nu}$, we obtain the field equation for $f(\mathcal{G})$ gravity as
\begin{equation}\label{5}
R_{\mu\nu}-\frac{1}{2}Rg_{\mu\nu}=T^{eff}_{\mu\nu},
\end{equation}
where $T^{eff}_{\mu\nu}$ is referred as
\begin{eqnarray}\nonumber\label{6}
&&T^{eff}_{\mu\nu}=\kappa^{2}(T_{\mu\nu}+E_{\mu\nu})-8[R_{\mu\rho\nu\sigma}+R_{\rho\nu}R_{\mu\nu}-R_{\rho\sigma}R_{\mu\nu}-g_{\mu\nu}R_{\rho\sigma}+R_{\mu\sigma}
\\ \label{determining1}&& g_{\nu\rho} +\frac{1}{2}\mathcal{R}(g_{\mu\nu}g_{\rho\sigma}-g_{\mu\sigma}g_{\nu\rho})]\nabla^{\rho}\nabla_{\sigma}f_{\mathcal{G}}+(\mathcal{G}f_{\mathcal{G}}-f)g_{\mu\nu}.
\end{eqnarray}
For our convenience, we use $\kappa^{2}$=$8\pi G$=1. It is worthy to notice that we have written the field equations in such a way that they look familiar with general relativity form. This approach seems interesting as it may provide all the matter components which are required to investigate the dark part of our universe. Thus it is expected that $f(\mathcal{G})$ theory of gravity may give fruitful results to understand the phenomenon of expansion of universe. In this way, the energy-momentum is not the usual one, but it plays effectively the similar role as in general relativity and also it accommodates dark matter. The first part of Eq. (\ref{6}) given in round brackets as $(T_{\mu\nu}+E_{\mu\nu})$ denotes the contribution of ordinary matter and second part of the sam equation in the large bracket justify the possibility of dark matter, as pointed out by many researchers \cite{a10a}. Here, the usual $T_{\mu\nu}$ is defined as
\begin{equation}\label{4}
T_{\mu\nu}=(\rho+p)u_{\mu}u_{\nu}+pg_{\mu\nu},
\end{equation}
where as $u_{\mu}$=$e^{\frac{\chi}{2}}\delta^{0}_{\alpha}$ is the four velocity vector. Just to avoid the confusion, it is to be noted that for our further analysis we will use the expression for the effective energy momentum tensor (\ref{6}).
 The charge term $E_{\mu\nu}$ is defined as
\begin{eqnarray}\label{7}
E_{\mu\nu}=\frac{g_{\mu\nu}}{2}[-F^{\mu\alpha}F_{\alpha\nu}+\frac{1}{4}\delta^{\mu}_{\nu}F^{\alpha\beta}F_{\alpha\beta}].
\end{eqnarray}
The Einstein-Maxwell tensor only contains the non-vanishing component $F^{01}$ and is defined as
\begin{eqnarray}\label{7a}
F^{01}=-F^{10}=\frac{q}{r^2}e^{-\frac{\chi+\xi}{2}},
\end{eqnarray}
where $q$ is the charge inside the spherical stellar system of the radial function $r$, given as
\begin{eqnarray}\label{7b}
q=4\pi \int_{0}^{r}\sigma \rho^2 e^{-\frac{\xi}{2}}d\rho,
\end{eqnarray}

We further consider Tolman-Kuchowicz space-time as
\begin{equation}\label{9}
\chi=wr^{2}+2\ln(x),~~~~~~~\xi=\ln(1+yr^{2}+zr^{4}),
\end{equation}
The metric potentials, $e^\chi$ and $e^\xi$, where $w$, $x$, $y$, and $z$ are arbitrary constants, proposed by Tolman and Kuchowicz, are chosen due to their singularity-free nature within the compact sphere interiors \cite{128,130}. Consequently, the chosen metric potentials are singularity free as shown in Fig. (1), which confirms the fulfillment of necessary conditions for these metric potentials.
Using all the above information, we obtain the following field equations as
\begin{eqnarray}\label{10}
\rho^{eff}+E^{2}= \frac{e^{-2\xi(r)}}{r^2}[-e^{\xi(r)}+e^{2\xi(r)}+r^2(\mathcal{G}f_\mathcal{G}-f)e^{2\xi(r)}+\xi^{'}(re^{\xi(r)}-4(e^{\xi(r)}-3)f_{\mathcal{G}^{'}})+8e^{\xi(r)}f_\mathcal{{G}^{''}}-8f_\mathcal{{G}^{''}}]+\rho,
\end{eqnarray}
\begin{eqnarray}\label{11}
p^{eff}-E^{2}=\frac{e^{-2\xi(r)}}{2r^2}[e^{\xi(r)}(2-e^{\xi(r)}(r^2f+2))+2r^2e^{2\xi(r)}\mathcal{G}f_\mathcal{G}+2\chi^{'}(re^{\xi(r)}-4(2e^{\xi(r)}-3)f_\mathcal{{G}^{'}})]+p,
\end{eqnarray}
\begin{equation}\label{11a}
\sigma=\frac{e^{-\frac{\xi}{2}}}{4\pi r^2}(r^2E)',
\end{equation}
whereas $\rho$ and $p$ denote the energy density and pressure component, respectively. There are three unknown functions in a system made up of two equations, namely $\rho^{eff}$, $p^{eff}$, and $E^{2}$. Here, $E^2$ represents the electric field and it is a component of the electromagnetic field. We further assume the internal stellar system by the simplification of MIT bag equation of state as
\begin{equation}\label{12}
p^{eff}=\frac{1}{3}\bigg[\rho^{eff}-4B_{g}\bigg],
\end{equation}
where $B_{g}$ is bag constant. By employing Eq. (\ref{12}), we obtain the expression for electric field $E^{2}$ as
\begin{equation}\label{13}
\begin{split}
E^{2}=&\frac{e^{-2\xi(r)}}{8r^2}\bigg[8e^{\xi(r)}-8e^{2\xi(r)}-16f^{'}\mathcal{G}^{''}+16e^{\xi(r)}f^{'}\mathcal{G}^{''}-8B_{g}r^{2}e^{2\xi(r)}+r^2f-4f_{\mathcal{G}}\mathcal{G}r^2-6pr^2+2r^2\rho-72\chi^{'}f^{'}\mathcal{G}^{'}\\ &+48e^{\xi(r)}\chi^{'}f^{'}\mathcal{G}^{'}-6re^{\xi(r)}\chi^{'}+24f^{'}\mathcal{G}^{'}\xi^{'}-8e^{\xi(r)}f^{'}\mathcal{G}^{'}\xi^{'}+2r\xi^{'}e^{\xi(r)}\bigg].
\end{split}
\end{equation}
\begin{figure}[!ht]
\includegraphics[scale=0.9]{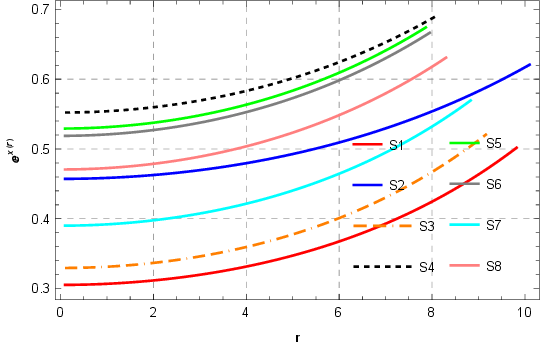}
\includegraphics[scale=0.9]{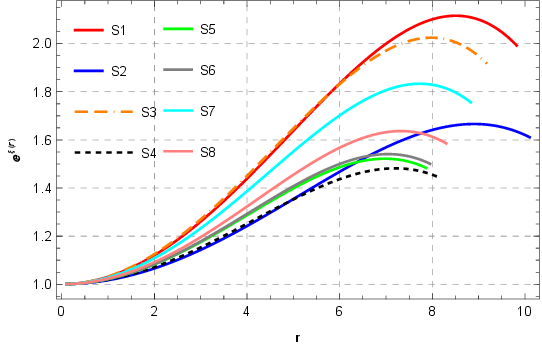}
\caption{Evolution of metric potentials $e^{\chi(r)}$ and $e^{\xi(r)}$ of candidate stars S1 (EXO 1785-248), S2 (PSR J1903+327), S3 (Cen X - 3), S4(SMC X - 4), S5 (LMC X - 4), S6 (Her X - 1), S7 (SAX J1808.4 - 3658) and S8 (4U 1538-52) against $r~(km)$.}\label{Fig.1}
\end{figure}
We consider the realistic $f(\mathcal{G})$ gravity model \cite{Bamba}, which reproduce the current cosmic acceleration, namely
\begin{eqnarray}\label{15}
f(\mathcal{G})=\tau\mathcal{G}^{n}(1+\psi\mathcal{G}^{m}),
\end{eqnarray}
where $\tau$, $\psi$ and $m$ are any constants, while $n>0$. The considered $f(\mathcal{G})$ model is considered worthwhile for the treatment of the finite time future singularities \cite{Noj3}.

\section{Matching with Exterior Metric}
In this section, we examine the matching conditions of spherically symmetric space-time with Bardeen as exterior geometry. There are many choices for the matching conditions but we consider Bardeen geometry for the present work. Therefore, the Bardeen space-time is defined as
\begin{equation}\label{16}
ds^{2} = D(r)dt^{2}-\frac{1}{D(r)}dr^{2}-r^{2}(d\theta^{2}+\sin^2\theta d\phi^{2}),
\end{equation}
where
\begin{equation}\label{20}
D(r)= 1-\frac{2M}{r}+\frac{3Mq^2}{r^3}+o\big(\frac{1}{r^5}\big).
\end{equation}
Here, $M$ is the gravitational mass and $Q$ represents the total charge. We employ the junction condition at $r=R$, which is given as
\begin{equation}\label{17}
g_{00}^{-}=g_{00}^{+},~~~~ g_{11}^{-}=g_{11}^{+},~~~~ \frac{\partial {g_{tt}^{-}}}{\partial{r}}=\frac{\partial {g_{tt}}^{+}}{\partial{r}},~~~~ \frac{\partial {g_{rr}^{-}}}{\partial{r}}=\frac{\partial {g_{rr}}^{+}}{\partial{r}},
\end{equation}
where (-) and (+) indicate the interior and the exterior solutions, respectively. After some manipulations, we get the values of unknown i.e, $w$, $x$, $y$ and $z$ as
\begin{equation}\nonumber\label{18}
y=\frac{M(-21q^2R^3+10R^5-4M(3q^{2}-2R^2)^2)}{2(3Mq^{2}R-2MR^3+R^{4})^{2}},~~~~~~~~~~~ w=\frac{2MR^2-9Mq^{2}}{2R^{2}(R^{3}-2MR^2+3Mq^{2})},
\end{equation}
\begin{equation}\nonumber\label{19}
z=\frac{M(15q^{2}R^3-6R^5+2M(3q^2-2R^2)^2)}{2R^4(3Mq^{2}-2MR^2+R^{3})^{2}},~~~~~~~~~~~ x=\sqrt{\frac{R^{3}-2MR^2+3Mq^{2}}{e^{wR^2}R^3}}.
\end{equation}
Now the numerical values of these constants can be calculated by using the observational data of various star candidates as given in TABLE. 1.
\begin{table}[h b t!]
\begin{center}
\caption{Unknown parameters of the stellar star candidates for $\tau=-55$,$\psi=20$, $n=2$ and $m=-1$}
\renewcommand{\arraystretch}{2}
\begin{tabular}{|c|c|c|c|c|c|c|c|}
\hline
\textbf{Star Models}& $M/M_\odot$ & $R(km)$ & $w(km)$ & $x(km)$ & $y(km)$ & $z(km)$ & $Z_s|_{r=R}$ \\
\hline
$\textbf{EXO 1785-248}(S1)$ & 1.667 & 9.82 &0.00299609 &0.676161 & 0.0168118 &-0.000106065 & 0.410573\\
\hline
$\textbf{PSR J1903+327}(S2)$ & 1.30  & 10.10 & 0.00514854 &0.552636 & 0.030855  &-0.000213185 & 0.265447 \\
\hline
$\textbf{Cen X - 3}(S3)$ & 1.49  & 9.178 &0.00544619  & 0.57413 & 0.032228  &-0.000253284 & 0.383882\\
\hline
$\textbf{SMC X - 4}(S4)$ & 0.85 & 8.1 &0.00484741  &0.624656 &0.027902  &-0.000233466 & 0.200896\\
\hline
$\textbf{LMC X - 4}(S5)$ & 0.87 & 7.866 & 0.00424821 & 0.686052 & 0.0237282 &-0.00022105 & 0.220077\\
\hline
$\textbf{Her X - 1}(S6)$ & 0.9 & 7.961 &0.00341256 &0.743066  &0.0185909  &-0.00017933 & 0.222254\\
\hline
$\textbf{SAX J1808.4 - 3658}(S7)$ & 1.29 & 8.831 & 0.00395975 & 0.720279 & 0.0217813 & -0.000219268 & 0.320792\\
\hline
$\textbf{4U 1538-52}(S8)$ & 1.04 & 8.301 & 0.00390841 & 0.727479 &0.0214324 &-0.000220053 & 0.250118\\
\hline
\end{tabular}
\renewcommand{\arraystretch}{1}
\end{center}
\end{table}

\section{PHYSICAL ANALYSIS FOR TOLMAN-KUCHOWICZ SPACE-TIME}
In this session, we study the different physical attributes of isotropic stellar structures.
Physical analysis of compact stars is essential for comprehensively understanding their structure, behavior, and evolution. This analysis involves investigating various physical characteristics such as energy density, pressure distribution, mass function, redshift function, and compactness factor. By examining these properties, researchers can discern crucial insights into the internal conditions, gravitational stability, and overall dynamics of compact stars. Additionally, studying the equation of state parameter allows for the characterization of the matter comprising these celestial bodies, shedding light on their thermal properties and composition. Furthermore, the fulfillment of energy conditions provides critical constraints on the energy-momentum tensor, ensuring the physical consistency and viability of theoretical models describing compact stars. Through rigorous physical analysis, scientists aim to unravel the mysteries surrounding compact stellar objects, advancing our knowledge of astrophysics and the cosmos as a whole. We analyze the physical viability of energy density, pressure component, energy conditions, EoS parameter, gradient of pressure and density, equilibrium condition, redshift function, mass-radius function, adiabatic index, and causality analysis.
\subsection{Energy Density and Pressure}
\begin{figure}[!ht]
\includegraphics[scale=0.9]{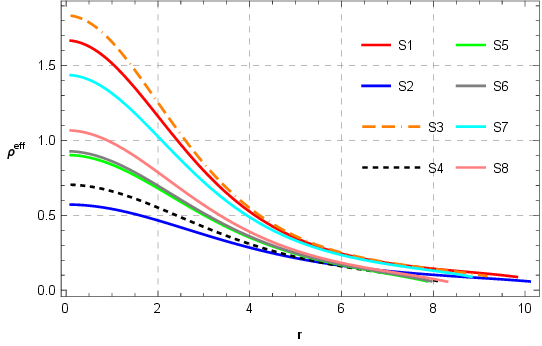}
\includegraphics[scale=0.9]{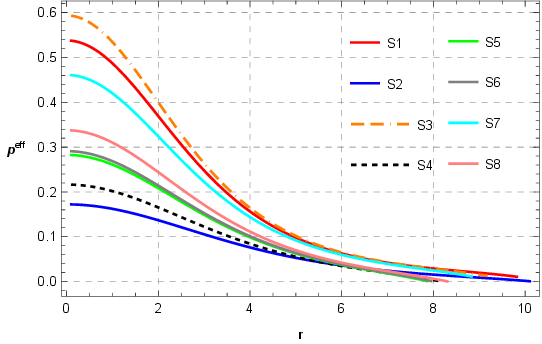}
\caption{Graphical behavior of $\rho^{eff}[MeV/fm^3]$, pressure $p^{eff}[MeV/fm^3]$ against $r~(km)$ from left to right.}\label{Fig.2}
\end{figure}

\begin{figure}[!ht]
\includegraphics[scale=0.9]{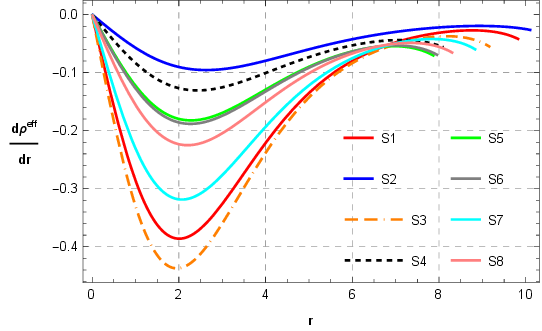}
\includegraphics[scale=0.9]{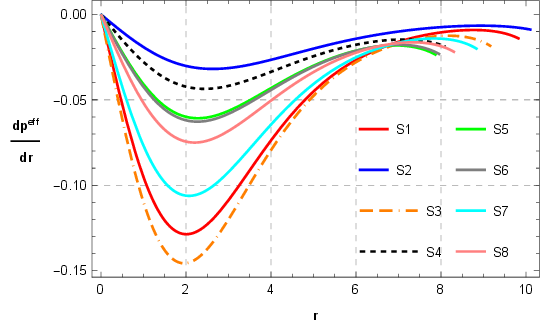}
\caption{Graphical behavior of gradients of $\rho^{eff}$ and $p^{eff}$  against $r~(km)$.}\label{Fig.3}
\end{figure}

The graphical representation of energy density and pressure components reach their peak at the core of the star, gradually decreasing towards the boundary of the star, as depicted in Fig. (\ref{Fig.2}). Additionally, the gradients of pressure and energy density demonstrate the negative trends as illustrated in Fig. (\ref{Fig.3}).

\subsection{Equation of State Parameter}
The Equation of State (EoS) parameter holds significant importance in discussions surrounding compact stars as it characterizes the relationship between pressure, density, and energy within these celestial bodies. By understanding the EoS, researchers can gain insights into the fundamental properties of matter under extreme conditions, such as those found within compact stars. This parameter provides crucial information about the internal structure, thermal properties, and stability of compact stars, aiding in the exploration of their formation, evolution, and observable phenomena. Furthermore, the EoS parameter plays a pivotal role in theoretical models and numerical simulations, guiding efforts to interpret observational data and validate theoretical predictions regarding the behavior and composition of compact stellar objects. Thus, the EoS parameter serves as a cornerstone for studying the physics of compact stars and advancing our understanding of the universe's most enigmatic entities. In our current study, we utilize
\begin{equation}\label{20}
\omega=\frac{p^{eff}}{\rho^{eff}}.
\end{equation}
It is to be noted that for the sake of simplicity to calculate $\omega$, we have considered $p$ and $\rho$ to be 1. We may choose any physical value of $p$ and $\rho$ to study how modified theories of gravity behaves. It's essential for the parameter $\omega$ to fall within the range of 0 and 1. In the context of compact stars, having the parameter $\omega$ between zero and one indicates that the pressure within the star is less than the energy density but not negligible. This range of values for $\omega$ encompasses a variety of physically plausible equations of state for the matter composing the interior of the compact star, including those involving degenerate fermions (like neutrons or quarks) or other exotic forms of matter. As observed in the left panel of Fig. (\ref{Fig.4}), the EoS parameter adheres to this specified range, confirming that the necessary conditions are met in this study.

\subsection{Causality Condition}
\begin{figure}[!ht]
\includegraphics[scale=0.9]{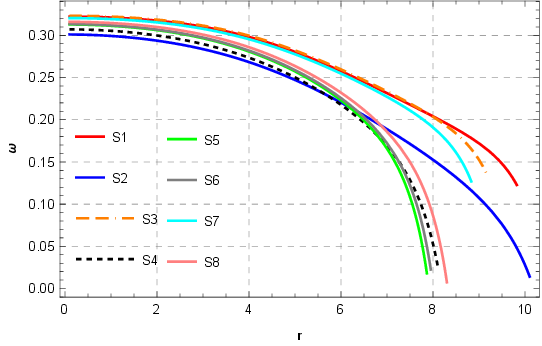}
\includegraphics[scale=0.9]{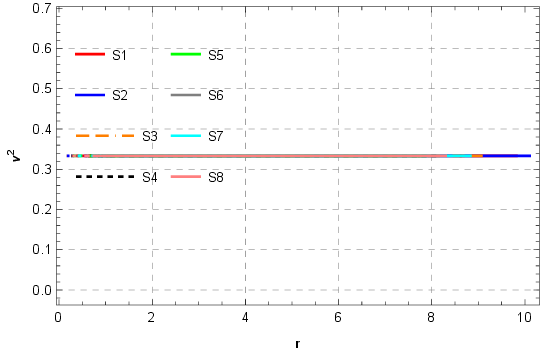}
\caption{Evaluation of $EoS$ parameter and sound speed against $r~(km)$.}\label{Fig.4}
\end{figure}

Ensuring the stability of stellar structures is pivotal for assessing the model's physical coherence. These celestial bodies hold a captivating allure, demonstrating resilience against external perturbations. Consequently, investigating stability has become paramount in understanding stellar evolution. In this study, we examine the stability of the considered star, adhering to Herrera's criteria \cite{141}. As per causal conditions, the sound speed should range between 0 and 1, implying that for stellar objects to be physically stable, $0\leq\upsilon^{2}\leq1$, where $\upsilon^2$ denotes the sound speed.
\begin{equation}\label{22}
\upsilon^{2}=\frac{dp^{eff}}{d\rho^{eff}}.
\end{equation}
The Herrera criterion plays a crucial role in determining the stability of compact stars by examining the behavior of the speed of sound within these stellar objects. Specifically, it states that the speed of sound should not exceed the speed of light within the star's interior to maintain stability against radial perturbations. This criterion provides valuable insights into the structural integrity and dynamical stability of compact stars, aiding in the identification of physical conditions conducive to stable stellar configurations. However, while extensively utilized in the framework of general relativity, the applicability of the Herrera criterion to alternative theories of gravity remains a subject of inquiry. The validity of this criterion across different gravitational frameworks requires careful investigation and empirical validation to ascertain its robustness and reliability in diverse astrophysical contexts. Further research and analysis are necessary to elucidate the criterion's suitability and limitations in the context of alternative gravitational theories, thereby enhancing our understanding of compact star stability.
The graphical evaluation of sound speed can be shown in the right panel of Fig. (\ref{Fig.4}), which clearly demonstrates that sound speed obey the given condition. The constancy of the sound speed across the radial coordinate, despite variations in the pressure and energy density of the effective fluid described by Eq. (18), can be understood through Eq. (\ref{12}), where the defined relation between pressure and energy density yields $\frac{dp^{eff}}{d\rho^{eff}}=\frac{1}{3}$. This observation indicates that the effective fluid behaves as an ultra relativistic fluid, leading to a consistent sound speed regardless of radial position.

\subsection{Tolman-Oppenheimer-Volkoff Equation}

\begin{figure}[!ht]
\includegraphics[scale=0.9]{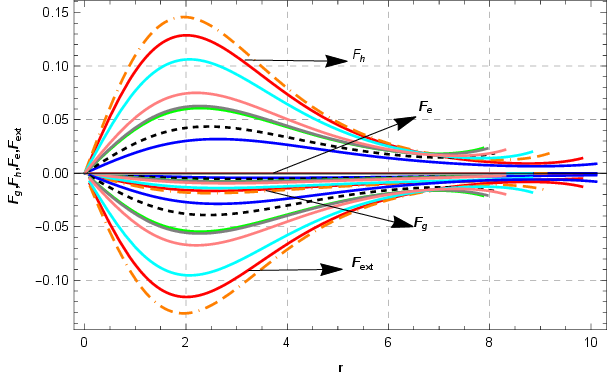}
\caption{\label{Fig.5} Behavior of equilibrium condition $[MeV/fm^3]$ against $r~(km)$.}
\end{figure}

The Tolman-Oppenheimer-Volkoff (TOV) equation holds pivotal importance in the study of compact stars, providing a fundamental framework for understanding their equilibrium structure. This differential equation, derived from general relativity, describes the balance between the gravitational inward pull and the outward pressure due to degeneracy of matter within a compact star. By solving the TOV equation, researchers can determine critical parameters such as the maximum mass and radius that a compact star can attain before collapsing into a black hole. This equation serves as a cornerstone in theoretical astrophysics, guiding investigations into the stability, composition, and observational characteristics of compact stellar objects. Through the application of the TOV equation, scientists aim to unravel the mysteries of extreme astrophysical phenomena and advance our understanding of the universe's most enigmatic entities. The TOV equation is described as
\begin{equation}\label{25}
-\frac{\chi^{'}(r)}{2}\bigg(\rho^{eff}+p^{eff}\bigg)+E\sigma(r)e^{\frac{\xi(r)}{2}}-\frac{dp^{eff}}{dr}+\frac{1}{7\pi}\bigg(\frac{d}{dr}(\rho^{eff}-p^{eff})\bigg)=0.
\end{equation}
It is to be notice that the TOV equation is a consequence of usual general relativity, as we have adopted an approach by writing field equations in such a way that they look familiar with general relativity form.
Eq. (\ref{25}) can be rewritten as sum of all forces as
\begin{equation}\label{30}
F_{g}\;+\;F_{e}\;+\;F_{h}\;+\;F_{ext}=\;0,
\end{equation}
where gravitational force ($F_{g}$=$-\frac{\chi^{'}(r)}{2}(\rho^{eff}+p^{eff})$), hydrostatic force ($F_{h}$=$-\frac{dp^{eff}}{dr}$), electric force ($F_{e}$= $E\sigma(r)e^{\frac{\xi(r)}{2}}$), and extra force ($F_{ext}$=$\frac{1}{7\pi}(\frac{d\rho^{eff}}{dr}-\frac{dp^{eff}}{dr})$). It can be observed from the Fig. (\ref{Fig.5}) that the graphical analysis of all the forces show the balancing behavior.


\subsection{Energy Conditions}
Energy conditions play a crucial role in discussing compact stars as they provide essential criteria for assessing the physical viability and stability of these celestial bodies. These conditions, including the Null Energy Condition (NEC), Weak Energy Condition (WEC), Dominant Energy Condition (DEC), and Strong Energy Condition (SEC), impose constraints on the energy-momentum tensor components within the framework of general relativity. By examining whether these conditions are satisfied, researchers can determine the behavior of matter and energy distribution within compact stars, helping to understand their structural integrity, gravitational stability, and overall physical consistency. Thus, energy conditions serve as fundamental guidelines for theoretical models and observational studies aimed at unraveling the mysteries of compact stellar objects. The energy conditions are defined as follows
\begin{figure}[!ht]
\includegraphics[scale=0.9]{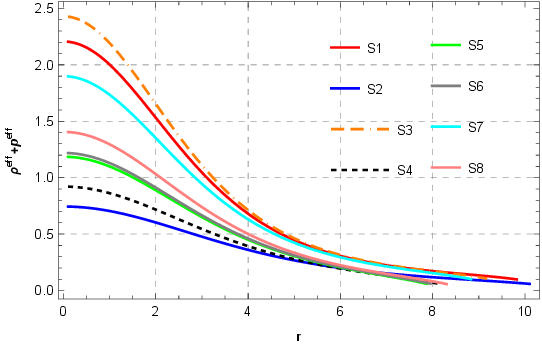}
\includegraphics[scale=0.9]{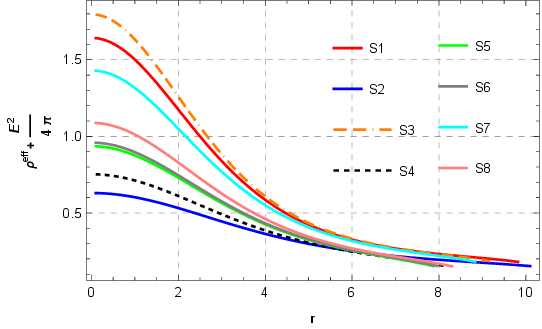}
\caption{\label{Fig.6} Graphical behavior of NEC $[MeV/fm^3]$ against $r~(km)$.}
\end{figure}
\begin{itemize}\label{26}
\item NEC: $\rho^{eff}+p^{eff}\geq0$,~~~~~ $\rho^{eff}+\frac{E^{2}}{4\pi}\geq0$.
\item WEC: $\rho^{eff}+\frac{E^{2}}{4\pi}\geq0$,~~~~~$\rho^{eff}+p^{eff}\geq0$,~~~~~$\rho+p^{eff}+\frac{E^{2}}{4\pi}\geq0$.
\item SEC: $\rho^{eff}+p^{eff}\geq0$,~~~~~ $\rho^{eff}+3p^{eff}+\frac{E^{2}}{4\pi}\geq0$.
\item DEC: $\rho^{eff}-p^{eff}+\frac{E^{2}}{4\pi}\geq0$.
\end{itemize}
It can be noticed from Figs. (\ref{Fig.6})-(\ref{Fig.8}) that all energy conditions are satisfied for our purposed model, which means that our considered stars are viable.

\subsection{Mass-Radius Relationship, Compactness Factor and Redshift Function }
The mass-radius relationship, compactness factor, and redshift function are key parameters used in the study of compact stars, providing valuable insights into their physical characteristics and gravitational properties.
\begin{figure}[!ht]
\includegraphics[scale=0.9]{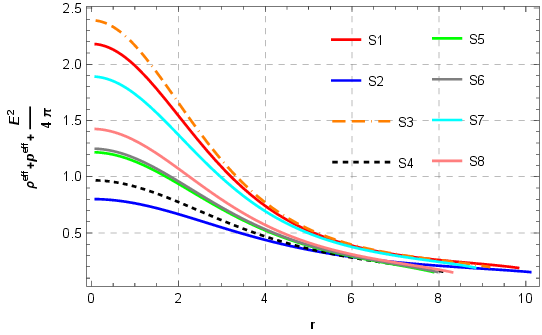}
\includegraphics[scale=0.9]{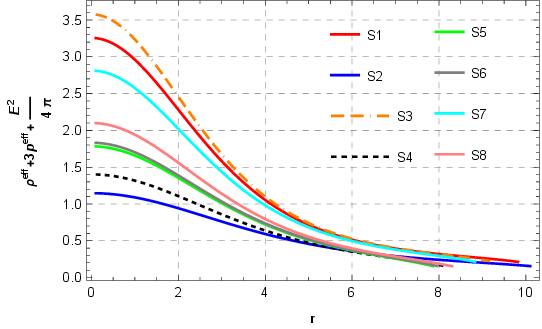}
\caption{\label{Fig.7} Graphical evaluation of WEC $[MeV/fm^3]$ and SEC $[MeV/fm^3]$ are shown against $r~(km)$.}
\end{figure}
 The mass-radius relationship quantifies the relationship between the mass and radius of a compact star, offering crucial information about its internal structure and density distribution. Understanding this relationship allows researchers to infer important details about the composition and equation of state of matter within the star. The mass-radius relation can be defined as
\begin{figure}[!ht]
\includegraphics[scale=0.9]{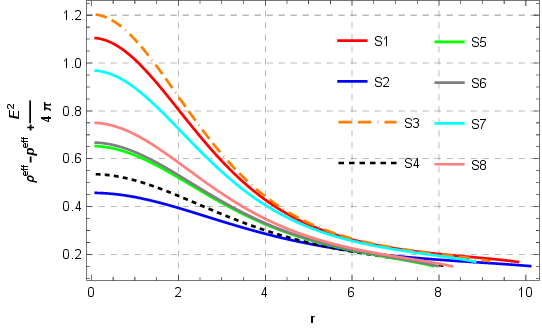}
\caption{\label{Fig.8} Graphical evaluation of DEC $[MeV/fm^3]$ against $r~(km)$.}
\end{figure}
\begin{equation}\label{31}
m(r)=\int^{R}_{0}4\pi\rho^{eff}r^{2}dr=\frac{1}{2}(1-e^{-\xi})R.
\end{equation}
The graphical analysis of mass-radius relations shows a positive and increasing behavior as seen in the left portion of Fig. (\ref{Fig.9}). The Compactness Factor, often denoted by the symbol $\mathcal{C}$, measures the degree of gravitational collapse within a compact star relative to its size. It is defined as the ratio of the star's mass to its radius, providing a measure of how densely packed the stellar material is. A high compactness factor indicates a strong gravitational field and a higher likelihood of exotic phenomena such as neutron degeneracy or black hole formation. The compactness element $U(r)$ is defined as
\begin{equation}\label{32}
U(r)=\frac{1}{r}\int^{R}_{0}4\pi\rho^{eff}r^{2}dr.
\end{equation}
The graphical behavior of compactness factor can be illustrated in the middle portion of Fig. (\ref{Fig.9}), which is minimum at the core and then becomes maximum, when we move towards boundary. The redshift function, on the other hand, refers to the shift in wavelength of photons emitted from the surface of a compact star due to gravitational effects. It is a consequence of general relativity and serves as a valuable tool for probing the gravitational field strength near the surface of the star. By measuring the surface redshift, astronomers can gain insights into the compactness and mass distribution of the star, as well as test the predictions of general relativity under extreme gravitational conditions. Furthermore, the surface redshift can be demonstrated as
\begin{equation}\label{33}
Z_{s}=(1-2U(r))^{\frac{-1}{2}}-1.
\end{equation}
The right panel of Fig. (\ref{Fig.9}) shows the graphical evaluation of surface redshift. It can also be noted that redshift function $Z_s$ at the surface $r=R$ has a specific value calculated for each star, which can be seen in Table I.
\begin{figure}[!ht]
\includegraphics[scale=0.8]{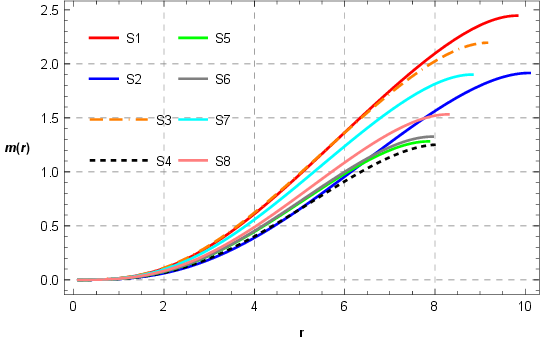}
\includegraphics[scale=0.8]{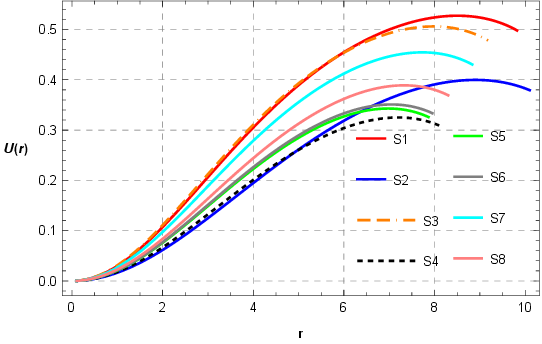}
\includegraphics[scale=0.8]{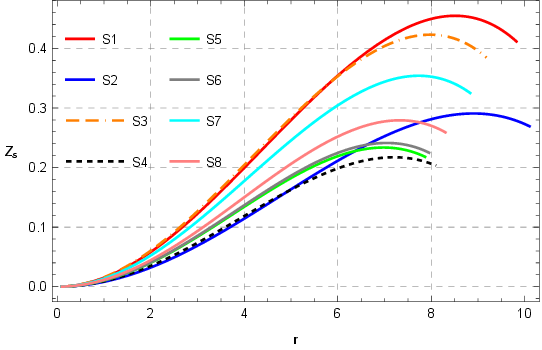}
\caption{\label{Fig.9} Graphical evaluation of $m(r)$, $U(r)$, and $Z_{s}$ are shown against $r~(km)$.}
\end{figure}
\section{Concluding Remarks}
The present study examine the charged isotropic compact objects in modified  $f(\mathcal{G})$ gravity by considering the Bardeen geometry as an exterior metric. In our current work, we use MIT bag EoS model i.e., $p^{eff}=\frac{1}{3}(\rho^{eff}-4B_{g})$. Furthermore, to get a constructive results of physical characteristics of stellar structure, we prefer Tolman-Kuchowicz metric along with isotropic stress-energy-momentum. Moreover, we employ the observational data of eight different stellar structures. Therefore, the brief qualitative analysis is itemized below:
\begin{itemize}

\item The metric potentials describe the geometry of space and time. Both metric
    potentials are at their lowest near the centre of the star and increase monotonically
     toward the surface. The graphical behaviour in Fig. (\ref{Fig.1}) clearly illustrates the stability
     and consistency of our metric potentials with the given requirements.

\item The graphical analysis reveals maximum pressure and density at the core of the star, gradually decreasing towards its boundary, as depicted in Fig. (\ref{Fig.2}). Additionally, Fig. (\ref{Fig.3}) illustrates negative and diminishing gradients of pressure and density.

\item  The evaluation of EoS ($\omega$) has decreasing behavior and lie in [0, 1], as demonstrated in the left portion of Fig. (\ref{Fig.4}).

\item The sound speed is frequently in the stability criteria i.e. $0\leq v^2\leq1$ as shown in the right panel of Fig. (\ref{Fig.4}), confirming the dynamical stability of our system.

\item Observing in Fig. (\ref{Fig.5}), it is evident that all forces exhibit a balanced behavior.

\item It's noteworthy that all energy constraints, including NEC, WEC, DEC, and SEC, are met by our proposed model, as depicted in Figs. (\ref{Fig.6})-(\ref{Fig.8}).

\item The graphical examination of additional physical characteristics such as mass function, surface redshift, and compactness factor is depicted in Fig. (\ref{Fig.9}).

\end{itemize}
Consequently, our considered $f(\mathcal{G})$ gravity model satisfies all the physical properties of stellar structure and our results also agree with \cite{125i}. Also, as seen in the case of equation of state parameter, the $f(\mathcal{G})$ gravity may provide some drastic results if we choose the predicted values for ordinary pressure and energy density of the star. Thus, $f(\mathcal{G})$ gravity may have some interesting results in deriving the mass-radius relations, moment of inertia, and tidal deformability of compact stars, and this is left for a future work.

\section*{Data Availability Statement}
The authors declare that the data supporting the findings of this study are available within the article.

\section*{Acknowledgement}
Adnan Malik acknowledges the Grant No. YS304023912 to support his Postdoctoral Fellowship at Zhejiang Normal University, China. Also, thankful to the anonymous referee whose constructive comments significantly contributed to the improvement of this paper.

\end{document}